\documentclass{amsart}
\usepackage{graphicx,tikz}
\usepgflibrary{arrows}
\usepackage[margin=1in]{geometry}
\begin{document}

\title[Gerrymandering metrics]{Gerrymandering metrics:\\ How to measure? What's the baseline?}
\author{Moon Duchin}
\maketitle

\vspace{-.2in}

\section{What is a districting plan?}

For the purposes of this brief discussion, I need to begin by specifying redistricting as a math problem in some way; that is,
by formalizing a districting plan as an appropriate kind of mathematical object.  I'll propose a way to do this that is completely 
uncontroversial:  start with the smallest units of population that are to be the building blocks of a plan---these might be 
units given by the Census, like  blocks, block groups, or tracts, or they might be units given by the state, like precincts or wards.%
\footnote{A districting plan normally won't go below the census block level,  because then population needs to be estimated. 
The units in which the election outcomes are recorded are called VTDs, or voting tabulation districts, 
which typically correspond to precincts or wards.
These are natural units to build a plan from if you want to study its partisan properties.}  
Represent those population units as nodes or vertices in a graph, and connect two of them if the units are geographically 
adjacent.  

For instance, here is a map of Wisconsin, and with it I have drawn a graph of its 1,409 census tracts.  You certainly 
can't see all the ones in Milwaukee by looking at this picture, because the graph is too dense there, which 
illustrates that plotting the graph
in this way (with the vertices at the centers of the tracts) also shows you where the population is clustered.

\begin{figure}[ht]
\includegraphics[width=2.6in]{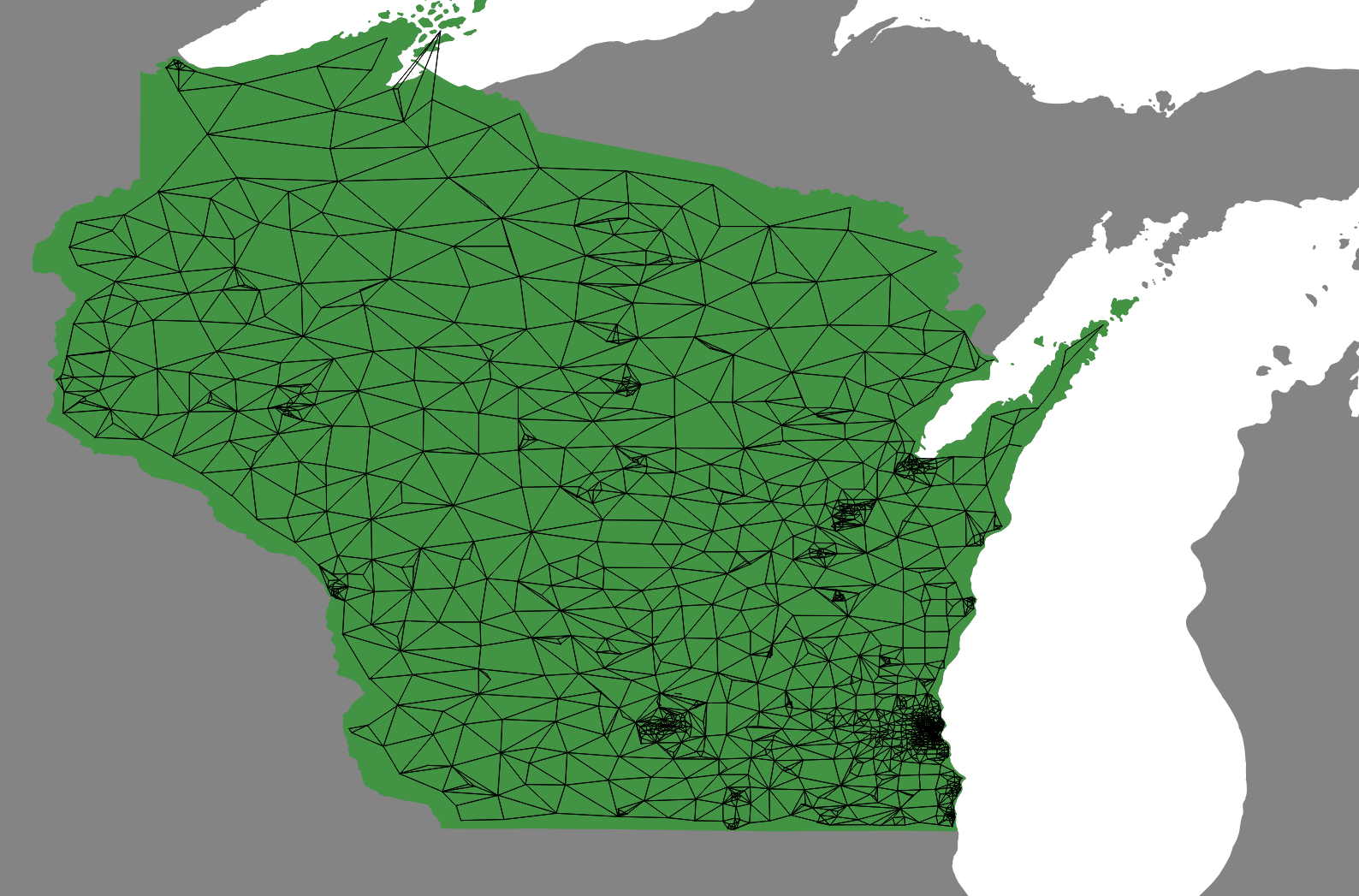} 
\raisebox{.1in}{\includegraphics[width=2.6in]{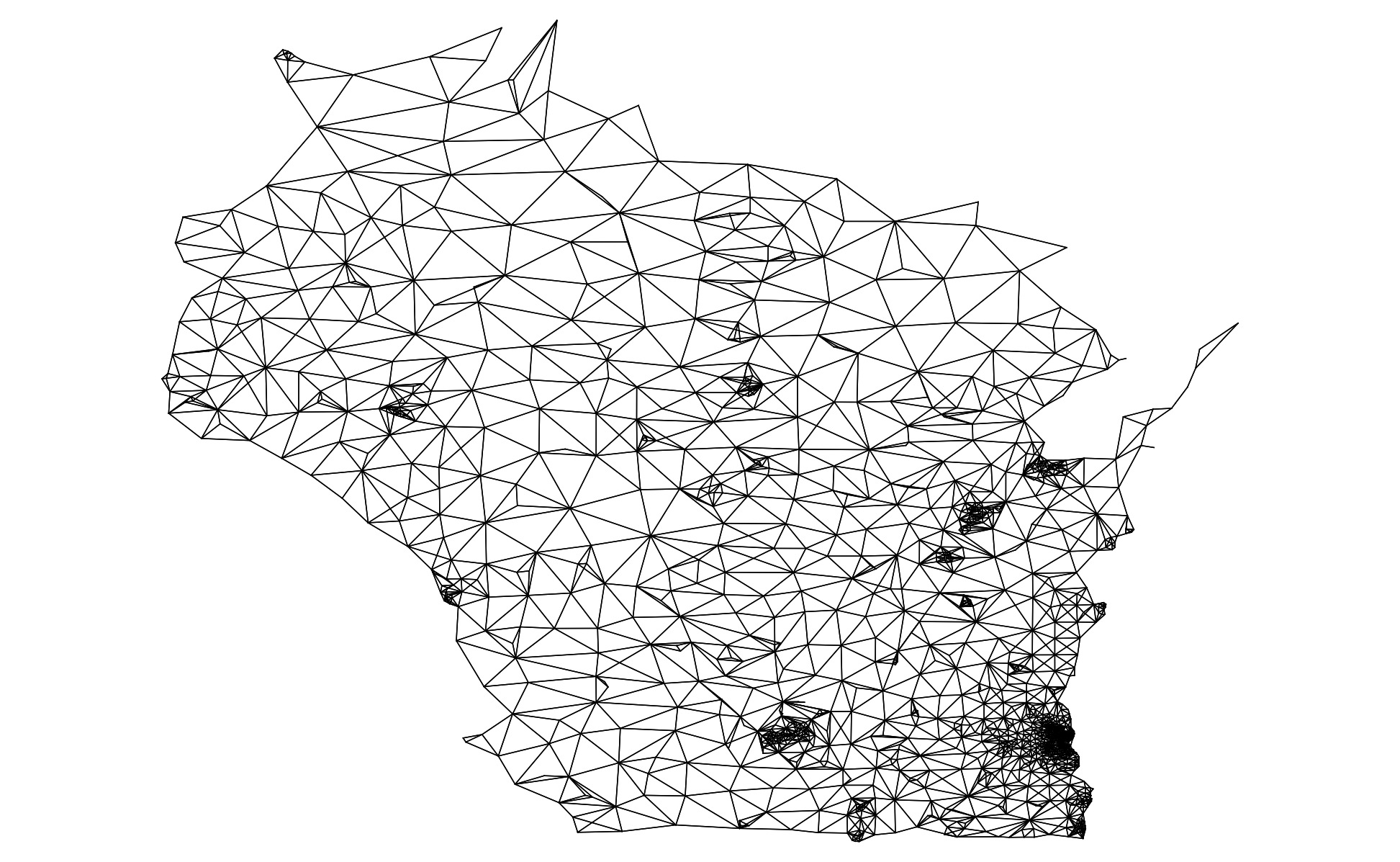} }
\caption{Wisconsin with the census-tract graph overlaid, and with the graph shown separately.}
\end{figure}

Armed with this, we can say that a (contiguous) {\em districting plan} is a partition of the vertices in the graph of a state's population
 into some number
of subsets called {\em districts}, such that each district induces a connected subgraph.\footnote{It is generally reasonable to also 
require that no district is wholly surrounded by another district.}
For instance, Wisconsin currently has eight congressional districts and 99 state assembly districts, so if they were made 
from census-tract units, then the former districts
would have 100-200 nodes each while the latter would have only 10-20 nodes.\footnote{That is, if the districts were made out 
of whole tracts.  In fact, they are typically made out of finer pieces, like precincts.}  

Our goal when we redistrict is to find a partition that meets a list of criteria. Some of those are universal and apply to the whole 
country, like having nearly equal populations in the districts and complying with the federal Voting Rights Act, and 
some are specified by states,  such as guidance about shapes of districts or about how much to allow the splitting of counties and cities.  Part of what makes redistricting so hard is that many of these rules are vague, and they often  represent
conflicting priorities.

The rest of this note will be devoted to outlining three intellectually distinct but not mutually exclusive strategies for 
measuring partisan gerrymandering.  The first two are {\bf only} suited for partisan gerrymandering, but the third is 
more flexible and can be used for other kinds of measurements of a plan, like racial bias or competitiveness.
For each approach we should track the norm and the baseline:  how does the metric correspond to a notion of fairness?  
and what's the basis of comparison against which a plan is assessed?  

\section{Partisan symmetry}

Partisan symmetry is a principle for districting plans that has been articulated and championed by Gary King, Bernie Grofman, 
Andrew Gelman, and several other prominent scholars.  (See \cite{GK} and its references.)
At its heart is a certain normative principle (or statement of 
how fair plans {\em should} behave):   
how one party performs with a certain vote share should be handled symmetrically if the other party received the same vote share.
That is, if Democrats got $60\%$ of the vote and, with that, won $65\%$ of the seats in the election, then if Republicans 
had earned $60\%$ of the vote, they too should have received $65\%$ of the seats.  

To visualize this, let's build a {\em seats-votes plot}.  On the $x$-axis we'll record $V$, the proportion of votes won by party $A$.
On the $y$-axis will be $S$, the proportion of seats in the electoral body won by $A$.  
So a single statewide election is represented by one point on this grid; for instance $(V,S)=(.6, .65)$.

The problem is that a single data point does not tell you enough to understand the properties of the districting plan as a
plan.
A standard way to extend this point to a curve is to use the model called {\em uniform partisan swing}:  look at the 
results district by district, and add/subtract the same number of percentage points to party $A$'s vote share in each district.
As you keep adding to $A$'s vote share, you eventually push the share past $50\%$ in some districts, causing those districts
to flip their winner from $B$ to $A$.  And as you subtract, you eventually push districts towards $B$.  Thus this method creates
a curve that is step-shaped, showing a monotonic increase in the proportion of seats for $A$ as the proportion of votes for $A$ 
rises.

\begin{figure}[ht]
\includegraphics[width=2.5in]{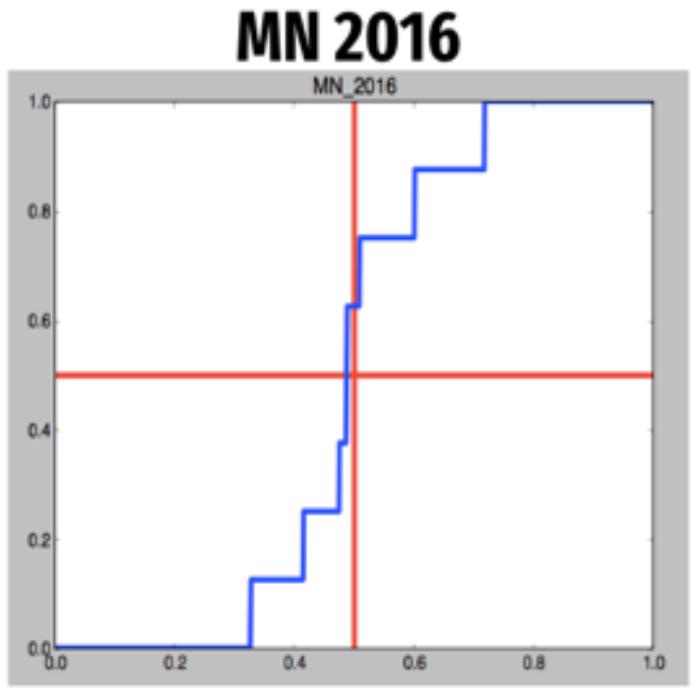} 
\includegraphics[width=2.5in]{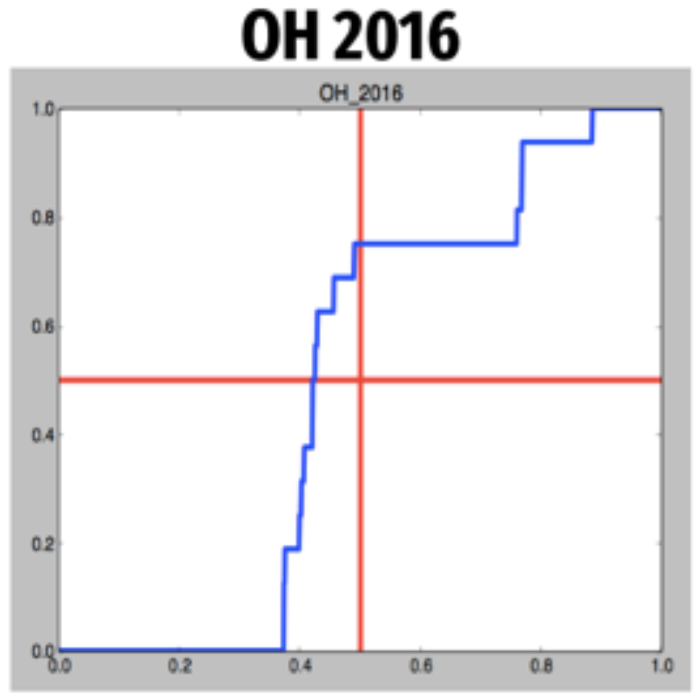} 
\caption{Seats-votes curves generated by uniform partisan swing from the Minnesota and Ohio congressional elections in 2016,
both presented from the Republican point of view.  The actual election outcomes from which the curves were derived are 
$(0.48, 0.38)$ for MN and $(0.58, 0.75)$ for OH.\label{SV}}
\end{figure}

In the figure, we see that the 
Minnesota election has a seats-votes curve that is very nearly symmetric about the center point $(.5,.5)$.  
On the other hand, Ohio's curve is very far from symmetric.  Rather, it looks like Ohio Republicans can secure 
$75\%$ of the congressional representation from just $50\%$ of the vote, and that just $42\%$ of the vote is enough for them
to take a majority of the congressional seats.  

A partisan symmetry standard would judge a plan to be more gerrymandered for producing a more asymmetrical seats-votes curve, 
flagging a plan if the asymmetry is sufficiently severe.
There are many ways that a mathematician could imagine using a bit of functional analysis to quantify the failure of symmetry,
but there are also a few elementary and easy-to-visualize scores:  for instance, look at how far the curve is from the 
center point $(.5,.5)$, either in vertical displacement or in horizontal displacement, on reasoning that any truly symmetric
plan must award each party half the seats if the vote is split exactly evenly.  Or compare one party's outcome 
with a given vote share to that for the other party if it had the same share.\footnote{If the seats-votes curve
is denoted $f(V)$, then this comparison amounts to $P(V)=\left|f(V)- \left[ 1- f(1-V) \right] \right|$.  It is natural, for instance, to evaluate
this at $V=V_0$, the actual vote share in a given election, but most authors don't commit to this.} 

In the (substantial) literature on partisan symmetry scores, you will sometimes see the vertical displacement 
called {\em partisan bias} (e.g., in \cite{EG}) and the horizontal displacement called the {\em mean/median score}.

\section{Efficiency gap}

Efficiency gap is a quite different idea about measuring partisan skew, both in its conceptual framing and in which plans 
it picks out as gerrymanders.  However, it does have a whiff of symmetry about it:  it begins with the normative principle 
that a plan is fair if the two parties ``waste" an equal number of votes. \cite{EG}

In order to parlay a certain given number of party-$A$ voters into the maximum possible number of seats for $A$, 
an extreme gerrymanderer would 
want to distribute $A$ voters {\em efficiently} through the districts:  you'd win as many districts as you can by narrow margins,
and you wouldn't put any of your voters at all in the districts won by the other side, because they're not contributing towards
your representation there.  By this logic, there are two ways for a party to waste votes.  On one hand, 
votes are wasted where there is an unnecessarily high winning margin---for this model, say every vote over $50\%$ is wasted.
On the other hand, all losing votes are wasted votes.  
Efficiency gap is the quantity given by the following simple expression:
just add up  the statewide wasted votes for party $A$ by summing over districts, subtract the statewide 
wasted votes for party $B$, and divide by the
total number of votes in the state.  Let's call this number $EG$.  Note that it is a signed score, and that 
$-.5\le EG \le .5$ by construction.\footnote{This is true because the total wasted votes in the state, and indeed
in each district, add up to half of the votes cast.}  
By the logic of the definition, a totally fair plan would have $EG=0$.
This score was first devised by political scientist Eric McGhee and was made into the centerpiece of a multi-pronged
legal test by McGhee and law professor Nick Stephanopoulos in their influential 2015 paper.  
For legislative races, they propose  $|EG|= .08$ as the threshold, past which a plan would be presumptively 
unconstitutional.  

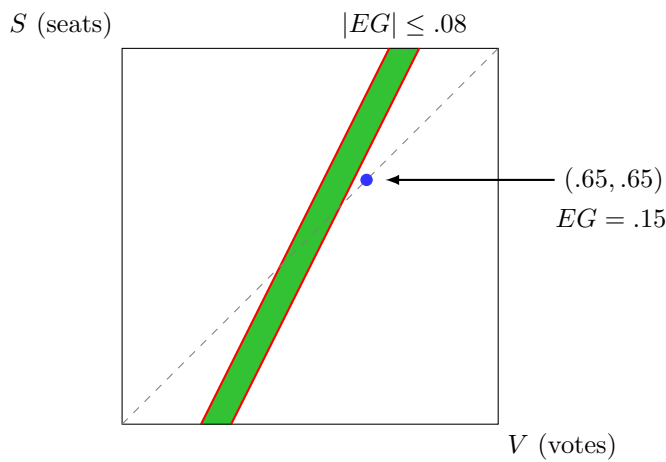
\begin{figure}[ht]
\begin{tikzpicture}[scale=5]

\filldraw [green!70!black,opacity=.8] (.21,0)--(.71,1)--(.79,1)--(.29,0)--cycle;

\draw (0,0) rectangle (1,1);
\node at (1,0) [below right] {$V$ (votes)};
\node at (0,1) [above left] {$S$ (seats)};
\node at (.75,1) [above] {$|EG|\le .08$};
\draw [thick,red] (.21,0)--(.71,1);
\draw [thick,red] (.29,0)--(.79,1);
\draw [gray,dashed] (0,0) -- (1,1) ;

\filldraw [blue!80] (.65,.65) circle (0.015);
\draw [-latex,thick] (1.15,.65)-- (.7,.65);
\node at (1.15,.65) [right] {$(.65,.65)$};
\node at (1.3,.55) {$EG=.15$};
\end{tikzpicture}
\caption{The region containing $(S,V)$ outcomes that pass the $EG$ test is shown in green.  The dashed line
is direct proportionality ($S=V$). \label{EGstrip} }
\end{figure}

Happily, this test is very easily represented on a seats-votes plot such as we introduced in the previous 
section.\footnote{Note that throughout this section we are leaning on the simplifying assumptions that all districts have equal turnout,
there are only two parties, and all races are contested by both sides.  These are varyingly realistic assumptions.}  
The permissibility zone (derived from the $EG$ formula and shown in Figure~\ref{EGstrip}) turns out to be a strip of slope two in the seats-votes space; any election that produces an outcome 
falling outside this zone is flagged as a gerrymander.  That the slope is two means that a certain ``seat bonus" is 
effectively prescribed for the winning side:  as the authors of the standard put it,
``To produce partisan fairness, in the sense of equal wasted votes for each party, the bonus should be a precisely twofold increase in seat share for a given increase in vote share." 
   This has the funny property that 
elections that produce directly proportional outcomes are often flagged as problematic.\footnote{As mentioned above,
$EG$ is proposed as one part of a multi-pronged legal test, so high $EG$ alone wouldn't doom a plan.  But it is obviously still relevant
to understand the systematic features of the score and the norms behind its construction.}
For instance, the point $(.65,.65)$ marked in the figure,
where a party has earned $65\%$ of the vote and converted it to $65\%$ of the seats, is seen as a gerrymander
in favor of the other side!  Quantitatively, that's because this case has $EG=.15$, far larger than the threshold.
Conceptually, it's because the party has received an inadequate seat bonus by the lights of the efficiency gap.

\section{Sampling}

Finally, I want to sketch a new approach to redistricting analysis that has started to crystallize only in the last five or so years.
It draws on a very well-established random walk sampling theory
whose growth has accelerated continuously
since its early development in the 1940s.\footnote{Diaconis's excellent survey \cite{diaconis}
reviews successful applications of MCMC in chemistry, physics, biology, statistics, group theory, and theoretical computer 
science.}   
The scientific details for the application to gerrymandering are still coalescing,
but the idea is incredibly promising and has  profound conceptual advantages that should cause it to fare well 
in courts.  This idea is to use algorithmic sampling to understand the space of all possible districting plans for a given state.

Remember our goal: we seek to split up a large, finite graph into some number of districts.  What you see in this picture is a very small graph being split up into four districts, represented by the different colors.  
First, we constrain the search space with requirements for valid plans, such as contiguity of the pieces, compactness of their shapes,
keeping population deviation under $1\%$, 
maintaining the current number of majority-minority districts, and so on.  (This will depend on the laws in place in the state we 
are studying.\footnote{The process of interpreting and 
operationalizing rules to create scores certainly bears scrutiny. A successful implementation will
have to demonstrate robustness of outcomes across choices made when scoring.})
A sampling algorithm takes a
random walk around the space of all valid partitions:  starting with a particular districting plan, flip units from district to district, 
thousands, millions, billions, or trillions of times.  

\begin{figure}[ht]
\includegraphics[width=4in]{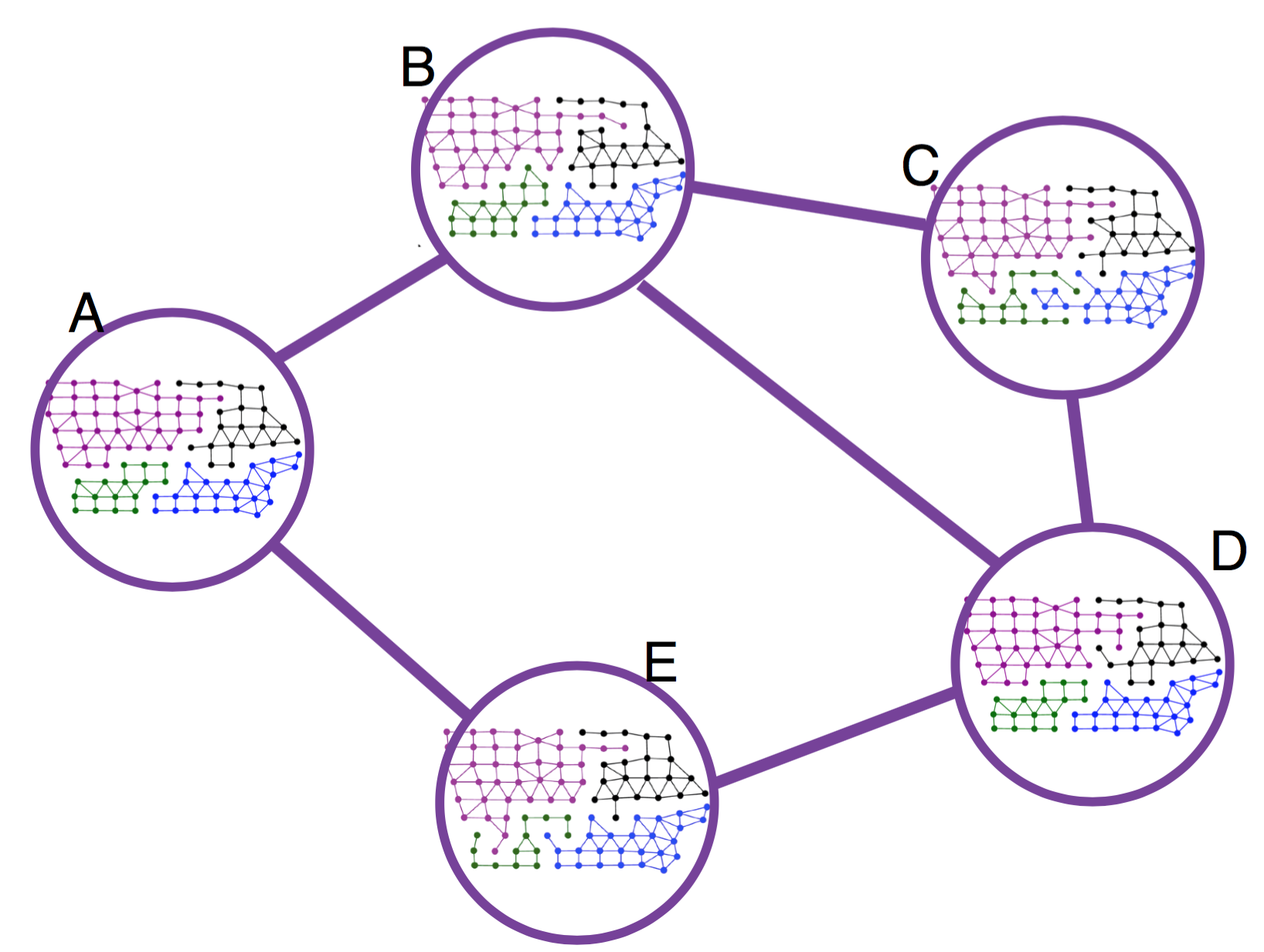}
\caption{This figure shows a tiny section of a search space of districting plans.  Two plans are adjacent here if a simple flip of a 
node from one district to another takes you from one plan to the other.  For instance, toggling one node between purple 
and black flips between plan $B$ to plan $C$.}
\end{figure}

Searching in in this way, such as with a leading method called Markov chain Monte Carlo, or MCMC, you can sample many 
thousands of maps from the chains produced by random flips.
Each one is a possible way that you could have drawn the district lines. 
Call this big collection of maps your {\em ensemble} of districting plans.

What can you do with a large and diverse ensemble of plans?
This finally gives us a good way to address the baseline problem that always looms over attempts to adjudicate
gerrymandering.  That is, it gives us a tool we can use to decide whether plans are skewed relative to other possible plans
with the same raw materials.
The norm undergirding the sampling standard is that districting plans should be constructed {\em as though} just by 
the stated principles.  

\begin{figure}[ht]
\includegraphics[width=4in]{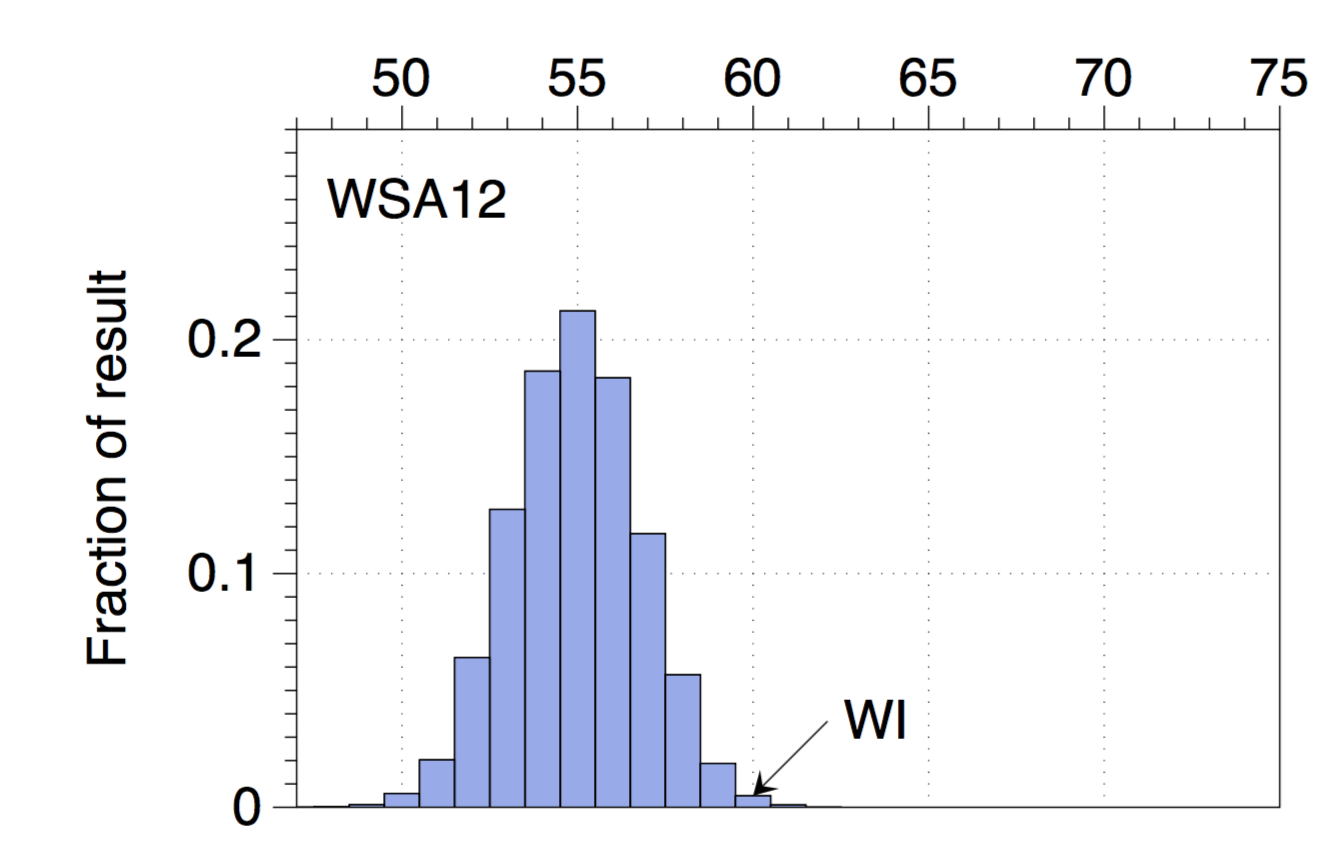}
\caption{On this plot, the $x$-axis is the number of seats won by Republicans out of 99 in the Wisconsin State Assembly,
and the $y$-axis shows how often each outcome occurred in the ensemble of $19,\! 184$ districting plans generated
by the MCMC algorithm from \cite{mattingly-WI}.  Given the actual vote pattern in the 2012 election, the plans could 
have produced a number of Republican seats anywhere from 49 to 61, with 55 seats being the most frequent outcome.
The legislature's plan (``Act 43") in fact produced 60 seats for Republicans, making it more extreme than $99.5\%$ of
alternatives.\label{histogram}}  
\end{figure}

The computer sampling methods could even be used to craft a new legal framework:  {\em Extreme outliers are impermissible}.  
(See Figure~\ref{histogram}.)
How extreme?  That would require some time and experience to determine, just as population deviation standards 
have taken some time to stabilize numerically in response to the corresponding legal framework {\em One person, one vote}.

The great strength of this method is that it is sensitive to the particularities, legal and demographic, of each state that 
it is used to analyze.  If a state has specific rules in its constitution or in state law---examples include North Carolina's 
``whole county provision," Wisconsin's quirky rules for district contiguity, Arizona's preference for competitive races,
incumbent protection in Kansas, 
Colorado's guidance to minimize the sum of the district perimeters---the sampling can be carried out subject
to those constraints or priorities.  And just as importantly, 
it addresses a major critique that can be leveled at both of the previous approaches:  why is it reasonable to prefer 
seats-votes symmetry, or to aim at equal vote wastage, when populations themselves are clustered in highly asymmetrical
ways? 
 For instance, imagine a state in which every household has 3 Republicans and 2 Democrats.  
(Of course this is highly unrealistic, but it's an extreme case of a state with a very uniform distribution of partisan preferences.)
Then no matter where you draw the lines, every single district will be $60\%$ Republican, which means Rs win $100\%$ 
of the seats, corresponding to the point $(.6,1)$ on the seats-votes plot.  One can easily verify that there is literally
{\em no plan at all} that doesn't have a sky-high partisan bias\footnote{In the simple model, the seats-votes curve is a 
step function with a big jump at $V=1/2$.  For more granularity, you could instead imagine a map in which one 
district has 39\% D and all others have percentages clustered around 41\%, also producing a high partisan bias
score for no very damning reason.  Compare MN-2016 from Figure~\ref{SV}.} 
or that gets the efficiency gap below $.3$.  
On the other hand, the sampling method will reveal an ensemble in which all plans are made up of 60--40 districts, 
and thus will show a particular plan with that composition 
to be completely typical and therefore permissible along partisan lines. It seems intuitively 
unreasonable for $60\%$ of the votes to earn all of the seats, but this method reveals that the political geography of 
this state demands it.

\section{Summary}

As mentioned earlier, these three approaches can be used in concert. For instance, one can use any evaluation
axis with a sampling ensemble, say efficiency gap (or mean-median score) instead of partisan outcome.  
So you can mix and match these approaches. Nonetheless, each has a different normative principle at its core and they 
would produce quite different redistricting outcomes if they were to be adopted at the center of a new legal framework.
Let's review some pros and cons.

For {\bf partisan symmetry}, it is really easy to make the case for fairness:  it sounds eminently reasonable that the two parties should
be treated the same by the system. 
Partisan symmetry uses up-to-date statistics and political science and has a lot of professional consensus behind it.
On the other hand, it has been critiqued by the court as too reliant on speculation and counterfactuals, 
mainly because of how it arrives at conclusions on how a plan would have performed at different vote levels.\footnote{As 
Justice Kennedy wrote:
``The existence or degree of asymmetry may in large part depend on conjecture about where possible vote-switchers will reside. Even assuming a court could choose reliably among different models of shifting voter preferences, we are wary of adopting a constitutional standard that invalidates a map based on unfair results that would occur in a hypothetical state of affairs."
LULAC v. Perry 126 S. Ct. 2594 (2006)}  
And it does not center on the question of how much advantage the line drawers have squeezed
from their power, because it lacks a baseline of how much symmetry a politically neutral agent could reasonably 
be expected to produce,
or even an agent who took symmetry as a goal.
Crucially, it is not at all clear that it is easy or even feasible to draw a map that will maintain partisan 
symmetry across several elections in a Census cycle.

An interesting and attractive feature of {\bf efficiency gap} is that it seems to derive, rather than prescribe, a permissible range 
in that seats/votes plot. It offers a single score and a standard threshold, and it is relatively easy to run.%
\footnote{The small print for efficiency gap:  you have to worry about which election data to use, and how to impute outcomes in 
uncontested races, which leaves a little bit of room for the possibility of dueling experts, but by and large it is a very manageable
standard.}  The creators of the $EG$ standard did about the best possible job of creating what the courts seemed to be
demanding:  a single judicially manageable  indicator of partisan gerrymandering.  The problem is that 
gerrymandering is a fundamentally multi-dimensional problem, so it is manifestly impossible to convert
that into a single number without a loss of information that is bound to produce many false positives or false negatives
for gerrymandering.  To illustrate the dimension issue, imagine that we are at your house and you ask me where I live.  
It is impossible to reasonably communicate the location of my house to you with a single number.  If you let me give you two numbers, I can give you latitude and longitude, say, but practically speaking just one number won't do.

Finally I have described  the {\bf sampling} approach and {\bf outlier analysis}, 
and I've argued that the strength of this approach is that it is sensitive to not only the law, as we've seen, but  also to the
political geography of each state---for instance, Wisconsin Democrats are densely arranged in Milwaukee proper, ringed by heavily Republican suburbs, but in Alaska Dems are spread throughout the rural parts of the state---which might have hard-to-measure effects on just how possible it is to split up the votes symmetrically or efficiently.  
Outlier analysis doesn't measure a districting plan against an all-purpose ideal, but against 
actual splittings of the state, holding the distribution of votes constant.   
In the next ten years, I expect to see explosive scientific progress on 
characterizing the geometry and topology of the space of districting plans, and on understanding the sampling distributions 
produced by our algorithms.


\vfill

\centerline{\fbox{
\begin{minipage}{5.9in}
{\em This note is a writeup of my presentation at the American Academy of Arts and Sciences during the 2062nd 
stated meeting on November 8, 2017, on the topic of {\em Redistricting and Representation}.  
Many thanks to the AAAS; the other panelists,
Gary King and Jamal Greene; and  the moderator, Judge Patti Saris.  Thanks also to Assaf Bar-Natan,
Mira Bernstein, Rebecca Willett, and the research team of Jonathan Mattingly, whose images are reproduced here with permission.
I am grateful to Mira Bernstein, Justin Levitt, and Laurie Paul for feedback.}
\end{minipage}
}}

\vfill

\end{document}